\documentclass[useAMS,usenatbib]{mn2e}
\usepackage{graphicx}
\title[L and T dwarfs in the Hyades \& Ursa Major moving groups]{L and T Dwarfs in the Hyades and Ursa Major moving groups}
\author[N.P.Bannister \& R.F.Jameson]{N.P.Bannister\thanks{E-mail:
npb@star.le.ac.uk} \& R.F.Jameson\\
Department of Physics \& Astronomy, University of Leicester, University Road, Leicester LE1 7RH, UK.\\}
\begin{document}

\date{Accepted xxxx xxxx xx. Received xxxx xxxx xx; in original form xxxx xxxx xx}

\pagerange{\pageref{firstpage}--\pageref{lastpage}} \pubyear{2002}

\maketitle

\label{firstpage}

\begin{abstract}
We have used the moving cluster method to identify three L dwarfs and one T dwarf in the Ursa Major/Sirius moving group (age 400 Myr). Five L dwarfs and two T dwarfs are found to belong to the Hyades moving group (age 625 Myr). These L and T dwarfs define 400- and 625- Myr empirical isochrones, assuming that they have the same age. Moving group membership does not guarantee coevality.
\end{abstract}


\section{Introduction}
Apart from a brief phase of lithium burning, brown dwarfs cool continuously. Thus any meaningful comparison with theory requires a knowledge of the age of the brown dwarf. For this reason much effort has been devoted to finding brown dwarfs in clusters whose age is known. The three closest clusters are the Hyades (d = 46 pc), Coma (d = 90 pc) and the Pleiades (d = 130 pc). The Hyades and Coma are old clusters with ages of 625 Myr \citep{Per:1998} and 500 Myr~\citep{Ode:1998} respectively, and were not thought to have any brown dwarfs. More recently~\citet{Mor:2003} have found 2 brown dwarfs in the Hyades and~\citet{Cas:2006} have found 13 brown dwarf candidates in the Coma cluster. The Pleiades (age 125 Myr) has some 50 known brown dwarfs~\citep{Jam:2002}, with some more recently discovered by \citet{Mor:2003}. Thus the nearest cluster with a significant number of known brown dwarfs is the Pleiades at a distance of 130 pc. This distance, together with the intrinsic faintness of brown dwarfs, naturally makes it difficult to study cluster brown dwarfs. By contrast, field brown dwarfs are close ($\sim 10 - 40$ pc) and easier to study but usually have unknown ages. However, some field star ages have been measured (see for example~\citet{Kir:2001} or~\citet{Burg:2006}). Field brown dwarfs are found by surveys such as 2MASS~\citep{Skr:1997}, DENIS~\citep{Eps:1997}, and the SDSS~\citep{Yor:2000}. A compilation of the known L and T dwarfs can be found in the L and T dwarf archive~\citep{Kir:2003}.

\begin{figure}
\vspace{-2cm}
\hspace{-0.20cm}
\resizebox{!}{11cm}
{\includegraphics{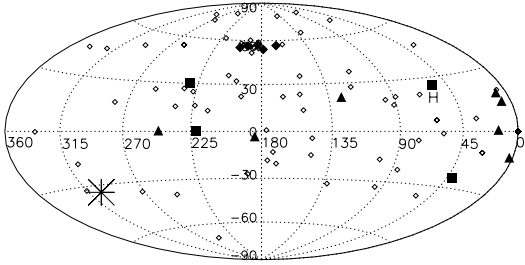}}
\vspace{-6.2cm}\caption{The location of the 77 UMSMG cluster members identified by Madsen et al. 2002 (open circles). The brightest members of the cluster, which along with another star make up the asterism of the ``Plough'' ($\beta, \gamma, \delta, \epsilon$ and $\zeta$ UMa) are shown as filled diamonds. The four new UMSMG members are shown as filled squares, and the location of the convergent point is indicated by the asterisk. Also shown on the map are the location of the seven new HMG members (filled triangles), and the location of the Hyades cluster (``H''). The coordinates are equatorial.}
\label{fig:map}
\end{figure}

One possible way of finding the ages of field brown dwarfs would be to see if they are members of a moving group. A moving group is a group of stars with the same velocity, magnitude and direction, and the same age (see \citet{Zuc:2004} for a recent review). One of the closest moving groups is the Ursa Major/Sirius moving group (hereafter UMSMG). The core of the moving group, possibly a bound cluster, is in the direction of Ursa Major. Indeed the stars of the ``Plough'' except $\alpha$ UMa are all members, as also is Sirius (see Fig.~\ref{fig:map}). Thus the Sun is actually inside the UMSMG. Group members can be found all around the sky, and may be very close: for example, Sirius is only 2.65 pc from the Sun. The age of the UMSMG has been determined as 300 Myr by \citet{Sod:1993}. More recently, \citet{Cas:2002} find 400 Myr while \citet{Kin:2003} find $500 \pm 100$ Myr for the group age. We will adopt an age of $400 \pm 100$ Myr. 

Since moving group stars have a common velocity they appear to be moving towards the same place in the sky; this is called the ``convergent point''. The UMSMG convergent point is located at $\alpha = 20^{h}18.83^{m},   \delta = -34^{\circ}25.8'$ (J2000 coordinates)~\citep{Mad:2002}. Thus if a field brown dwarf has a proper motion directed towards the UMSMG convergent point, it is a potential member of the UMSMG. This, coupled with two distance tests (see below), allows us to identify members with considerable confidence.

The Hyades is discussed in a thorough paper by \citet{Per:1998}. The cluster lies at a distance of d = 46 pc, and has an extent in the sky of approximately $20^{\circ}$;~\citet{Mad:2002} give the position of the cluster centroid as $\alpha = 4^{h}26^{m}, \delta = +16^{\circ}54'$. The Hyades is known to be deficient in low mass members~\citep{Giz:1999}. These have probably evapourated from the cluster. Indeed,~\citet{Che:1998} have identified escaped Hyads and these may be thought of as part of the Hyades Moving Group (HMG). The convergent point of the Hyades is located at $\alpha = 6^{h}29.48^{m}, \delta = -6^{\circ}53.4'$ and their total space velocity is 46 km/sec~\citep{Mad:2002}. The most recent and generally quoted Hyades age is $625 \pm 50$ Myr by~\citet{Per:1998}, and we will adopt this age.

\section[]{Identifying Group Members}
For the 70 members of the L and T Dwarf Archive \citep{Kir:2003} (available on the worldwide web at www.dwarfarchives.org) with a measured proper motion, we first calculate the angular distance, $D$, of the dwarf from the UMSMG or HMG convergent point, where $D$ is given by

\begin{equation}
   \cos D = \sin \delta \sin \delta_{cp} + \cos \delta \cos \delta_{cp} \cos DA.
\end{equation}
Here $\delta$ and $\delta_{cp}$ are the dwarf declination and convergent point declination, and $DA$ is the difference of their respective Right Ascensions.

Next we find the direction, $\theta$, from north of the convergent point, where

\begin{equation}
\cos \theta = \frac{\sin \delta_{cp} - \sin \delta \cos D}{\cos \delta \sin D}.
\end{equation}

A group member should have a proper motion direction equal to $\theta$. However, there is some velocity dispersion amongst group members, otherwise all group members would appear to be very close together, as if in a bound cluster. For the UMSMG the velocity $v$ is 17.98 km s$^{-1}$ and $\sigma_{v} = 2.82$ km s$^{-1}$~\citep{Mad:2002}. We adopt the same $\sigma_{v}$ for the Hyades' recent escapers, even though this value is considerably less than the $\sqrt{3.6^2+3.2^2+5.2^2}=7.09$ km s$^{-1}$ given by~\citet{Che:1999}. Thus we impose the same membership conditions for both the UMSMG and the HMG.

We find members have a proper motion direction within $\sim 13^{\circ}$ of $\theta$. This corresponds to $1.5\sigma_{v}$, or 87\% completeness, which seems reasonable. This constraint is our first criterion for membership, and the random chance of passing this first test is clearly $4\times 2\times 13 / 360 = 0.28$. The extra factor of 4 is because proper motion directions are not randomly orientated (see Section 7). 



It may readily be shown \citep{Car:1996} that for a moving group the distance $d_{mc}$ (in parsecs) of any member is given by
\begin{equation}
d_{mc}=\frac{v \sin D}{4.74 \mu},
\end{equation}
where $\mu$ is the proper motion in arcsecs per year. If the star is not a moving group member then the above formula does not apply. Our second test is to compare this moving cluster distance to the distance measured by parallax, $d_{p}$. Once again 1.5 times the velocity dispersion leading to a 28\% error compared to the parallax distance seems to cover all the members we find. As in the first test, we estimate the random chance of a star passing this test. Using the 70 dwarfs with parallaxes, minus the 4 dwarfs which we ultimately identify as UMSMG members (as discussed in the next section), we calculate $d_{mc} / d_{p}$, and find that 9 dwarfs have $0.72 < d_{mc} / d_{p} < 1.28$ i.e. within 1.5 $\sigma_{v}$, or 28\%. If 9 out of 66 dwarfs pass this test by chance, the probability is $9/66 = 0.14$. A similar test for the HMG yields $14/63 = 0.22$. We adopt this higher probability for both UMSMG and HMG to avoid over-estimating the significance of the test outcomes.

Finally we calculate the absolute magnitude at any wavelength using the parallax, and our last test is to place the objects in a colour - absolute magnitude diagram. This third check requires that the object lies in a ``correct'' or sensible position in the colour-magnitude diagram. By that we mean that there is some evident sequence. We do not require that the objects fit the theoretical isochrones (see point (v) under Section 6).

The entire L dwarf sequence for the 70 field stars is approximately 3.5 magnitudes wide, a factor 25 in intensity. Allowing for binaries, an isochrone can vary in intensity at any colour by a factor 2. This gives $2/25 = 0.08$ as the random chance of passing the third test.

Thus the total probability of passing all three independent tests by chance is $0.28 \times 0.22 \times 0.08 = 0.50\%$, suggesting that passing all three tests gives 99.50\% confidence of membership.

The dwarf archive has some 459 entries but unfortunately only 70 of these have measured proper motions. Those with proper motions also have accurate parallaxes.

\section{L and T dwarfs in the UMSMG}
Of the 70 objects in the archives with proper motions we find 4 to be members of the UMSMG. Three are L dwarfs and there is one T dwarf. Table~\ref{tab:member_basic} lists their spectral type, magnitude, and distance as determined from the moving cluster and parallax methods. Also in Table~\ref{tab:member_basic} we give $\Delta\theta$, the difference between the convergent point direction and measured proper motion direction. As mentioned above, due to velocity dispersion and errors in the moving group we do not expect $\Delta \theta$ to be zero or $d_{mc} / d_{p} = 1$. Velocity dispersion dominates over measurement errors. As can be seen from Table~\ref{tab:member_basic}, $\Delta \theta$ varies from $1.6^{\circ}$ to $13.5^{\circ}$ and $d_{mc} / d_{p}$ differs from unity by 1\% to 18\%. With these two parameters the group members effectively pick themselves. Thus if $\Delta \theta$ is allowed to increase above $13^{\circ}$ to say $25^{\circ}$, no candidates have $d_{mc} / d_{p}$ close to unity.

However, in the interests of scientific integrity, we point out that a fifth star, 2M1228-15, passed the first two tests and apparently passed the third. However 2M1228-15 is a known near-equal mass binary~\citep{Bra:2004}. Thus its two components lie $\sim ~0.75$ mag below their combined magnitude and do not fit the UMSMG sequence, hence failing the third test. This does not fit very well with our estimate of $\sim 3$\% chance of passing the first two tests by random chance.

\section{L and T dwarfs in the HMG}
For the HMG we find that 5 L dwarfs and 2 T dwarfs pass all three tests. Again relaxing the constraints on $\Delta\theta$ and $d_{mc}/d{p}$ would find no further members. However, 2M0205-1159 (also known as DENIS-P J020529.0-115925) was found to be a binary by ~\citet{Koe:1999} who measured $K$-band flux ratios of $1.00 \pm 0.26$ and $0.99 \pm 0.08$. More recently~\citet{Bou:2005} claim it is a triple system with I magnitudes of 17.30, 18.38 and 18.80, and spectral types L5.5, L8 and T0. These parameters suggest that the primary would have more than half of the K flux and so should not be moved down 0.75 mag in the colour-magnitude diagram. We mark 2M 0205-1159 with a downward-pointing arrow in Figs~\ref{fig:mk_jk} and~\ref{fig:mk_hk} (later) and regard its membership of the HMG as uncertain.

\begin{figure}
\vspace{-1cm}
\hspace{-0.75cm}
\resizebox{!}{13cm}
{\includegraphics{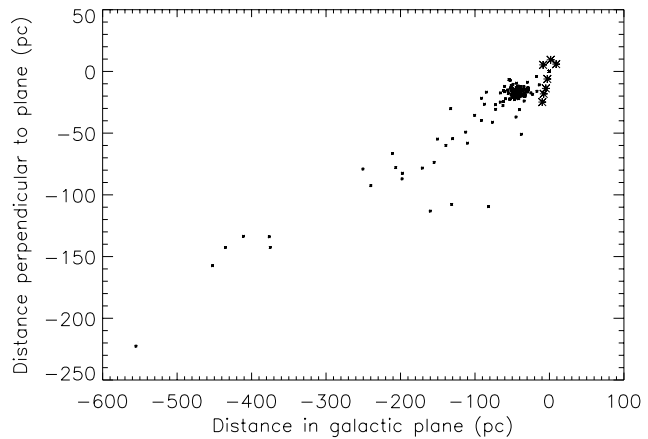}}
\vspace{-6cm}\caption{Location of the Hyades moving group members (points) and seven group members identified in this work (asterisks). Distances are in parsecs from the sun, in the radial sense (i.e. the component of distance parallel to the plane of the galaxy, indicated in the X axis) and perpendicular to the plane (Y axis). The newly identified members appear to follow the general distribution of the cluster; this agreement is also observed in the X-Y and Y-Z planes.}
\label{fig:hyades_stream}
\end{figure}

Figure~\ref{fig:hyades_stream} shows the location of the HMG group in galactic radial direction (X axis) and perpendicular to the plane (Y axis). The sun is at (0,0). The cluster is obvious and most of the moving group members form a stream in approximately the galactic anticentre direction, but with a few in front of the cluster. The 7 dwarf members are shown as asterisks. The uncertain binary member, 2M0205, has the most negative distance perpendicular to the plane, and is thus at the extreme end of the group. This might be considered as further evidence of its non-membership. The stream towards the galactic centre looks very similar to, but shorter than, that away from the galactic centre. The reason for the short length of the forward stream is no doubt because most surveys for Hyads have been conducted in the general direction of the Hyades.

\begin{table*}
\caption{Summay of the four brown dwarf members of the Ursa Major moving group and seven brown dwarf members of the Hyades moving group identified in this work. IR Spectral types and magnitude are listed, along with distances estimated from parallax measurements ($d_p$) and the moving cluster method ($d_{mc}$), and the difference between predicted and observed proper motion direction ($\Delta\theta$).}
\label{tab:member_basic}
\begin{center}
\begin{tabular}{lccccrrr}\hline
2MASS ID & IR Spectral &  m$_{J}$ & m$_{H}$ & m$_{K}$ & d$_{p}$ & d$_{mc}$ & $\Delta\theta$ \\ 
 & type & & & &  (pc) & (pc) & ($^{\circ}$)  \\
\hline
\textbf{UMSMG}\\
2M J02431371-2453298 	&T6 		& 15.381 $\pm$ 0.050 	& 15.137 $\pm$ 0.109 	& 15.216 $\pm$ 0.168 	& 10.7  $\pm$ 0.4	& 10.6 $\pm$ 0.5 & 1.64  	\\
2M J03454316+2540233	&L1  $\pm$ 1 	& 13.997 $\pm$	0.027	& 13.211 $\pm$ 0.030	& 12.672 $\pm$ 0.024 	& 27.0  $\pm$ 0.4	& 31.7 $\pm$ 1.2 & 5.62  	\\ 
2M J14460061+0024519	& L6		& 15.894 $\pm$ 0.082	& 14.514 $\pm$ 0.035 	& 13.935 $\pm$ 0.053	& 22.0  $\pm$ 1.5	& 19.7 $\pm$ 1.5 & 13.54 	\\
2M J15232263+3014562$^{*}$	& L8		& 16.056 $\pm$ 0.099	& 14.928 $\pm$ 0.081 	& 14.348 $\pm$  0.067 	& 18.6  $\pm$ 0.4	& 17.1 $\pm$ 1.1 & 11.70 \\
\\
\textbf{HMG}\\
2M J16241436+0029158	& T6 		& 15.494 $\pm$ 0.054 	& 15.524 $\pm$ 0.100 & 15.518 $\pm$ 0.000 	& 11.0 $\pm$ 0.1	& 13.9$\pm$ 0.2 	& 14.95 	\\
2M J0036159+182110 	& L4 $\pm$ 1 	& 12.466 $\pm$ 0.027	& 11.588 $\pm$ 0.029 & 11.058 $\pm$ 0.021 	& 8.8	$\pm$ 0.1	& 10.8	$\pm$ 0.1 	& 1.48 		\\ 
2M J00325937+1410371	& L8		& 16.830	$\pm$ 0.169	& 15.648 $\pm$ 0.142 & 14.946 $\pm$ 0.109 	& 33.2	$\pm$ 6.9	& 35.4	$\pm$ 1.2 	& 0.98 		\\ 
2M J0205293-115930		& L5.5 $\pm$ 2 	& 14.587 $\pm$ 0.030	& 13.568 $\pm$ 0.037 & 12.998 $\pm$ 0.030	& 19.8	$\pm$ 0.6	& 20.8 $\pm$	0.2 	& 5.21 		\\
2M J01075242+0041563	& L5.5		& 15.824 $\pm$ 0.058 	& 14.512 $\pm$ 0.039 & 13.709 $\pm$ 0.044 	& 15.6	$\pm$ 1.2	& 15.2	$\pm$ 0.3 	& 1.10 		\\
2M J1217110-031113 	& T7.5 		& 15.860 $\pm$ 0.061	& 15.748 $\pm$ 0.119 & 15.887 $\pm$ 0.000 		& 11.0	$\pm$ 0.3	& 9.2	$\pm$ 0.1 	& 2.86 		\\
2M J0825196+211552 	& L6 		& 15.100 $\pm$ 0.034	& 13.792 $\pm$ 0.032 & 13.028 $\pm$ 0.026 	& 10.7 $\pm$ 0.1	& 8.7 $\pm$ 0.1 		& 6.38 	\\
\hline
{\footnotesize $^{*}$Also known as Gl584C}

\end{tabular}\\
\end{center}
\end{table*}

\section{Notes on individual stars}
{\em 2M0243.}~\citet{Burg:2006} list the effective temperature of this star as $1040 \leq T_{eff} \leq 1100$ K, with log $g$ in the range 4.8-5.1 and an age of between 0.4 - 1.7 Gyr. This age range just fits to our adopted age for the UMSMG.\\
{\em 2M1523.} Also known as Gl584C, this star, which we include as a member of the UMSMG, is considered extensively by~\citet{Kir:2001} who estimate its age to be between 1.0 and 2.5 Gyr. This age is the average of several methods which have a total range of 0.3 to 2.5 Gyr, and thus encompasses the UMSMG age of 400 Myr.\\
{\em 2M1624.}~\citet{Burg:2006} list the effective temperature of this star as $980 \leq T_{eff} \leq 1040$ K, with log $g$ in the range 5.3-5.4 and an age of between 4.3 - 5.8 Gyr. This age is in clear conflict with the HMG age of 625 Myr. The method used by Burgasser et al is to find $g$ and $T_e$ from spectral indices, and compare these values with the model of~\citet{Bur:1997} which yields masses and ages directly. They also use measured luminosities to obtain masses and radii and then the models again to find the ages. This alternative method gives an age of 0.6 to 10 Gyr, just consistent with the Hyades age.\\
{\em 2M0036.}~\citet{Ber:2005} present a study of the magnetic properties and summarise current research on this object, citing $T_{eff} = 1923^{+193}_{-153}$ K~\citep{Vrb:2004}, log $g \approx 5.4$~\citep{Sch:2001}, and an inferred age of at least 1 Gyr from the work of~\citet{Bur:2001}.\\
{\em 2M0205.} This object is a known binary~\citep{Koe:1999};~\citet{Bou:2003} assume an age ``greater 0.5 Gyr'' but in later work,~\citet{Bou:2005} present evidence to suggest that 2M0205 is possibly a triple system and they assume an age of between 1 and 10 Gyr.\\
{\em 2M1217.}~\citet{Burg:2003} suggest the possibility of a faint companion to this object in Hubble WFPC-2 data. However, the putative companion is close to the detection limits of the image.\\

\section{Discussion}

\begin{figure}
\vspace{0cm}
\hspace{-0.5cm}
\resizebox{!}{6.9cm}
{\includegraphics{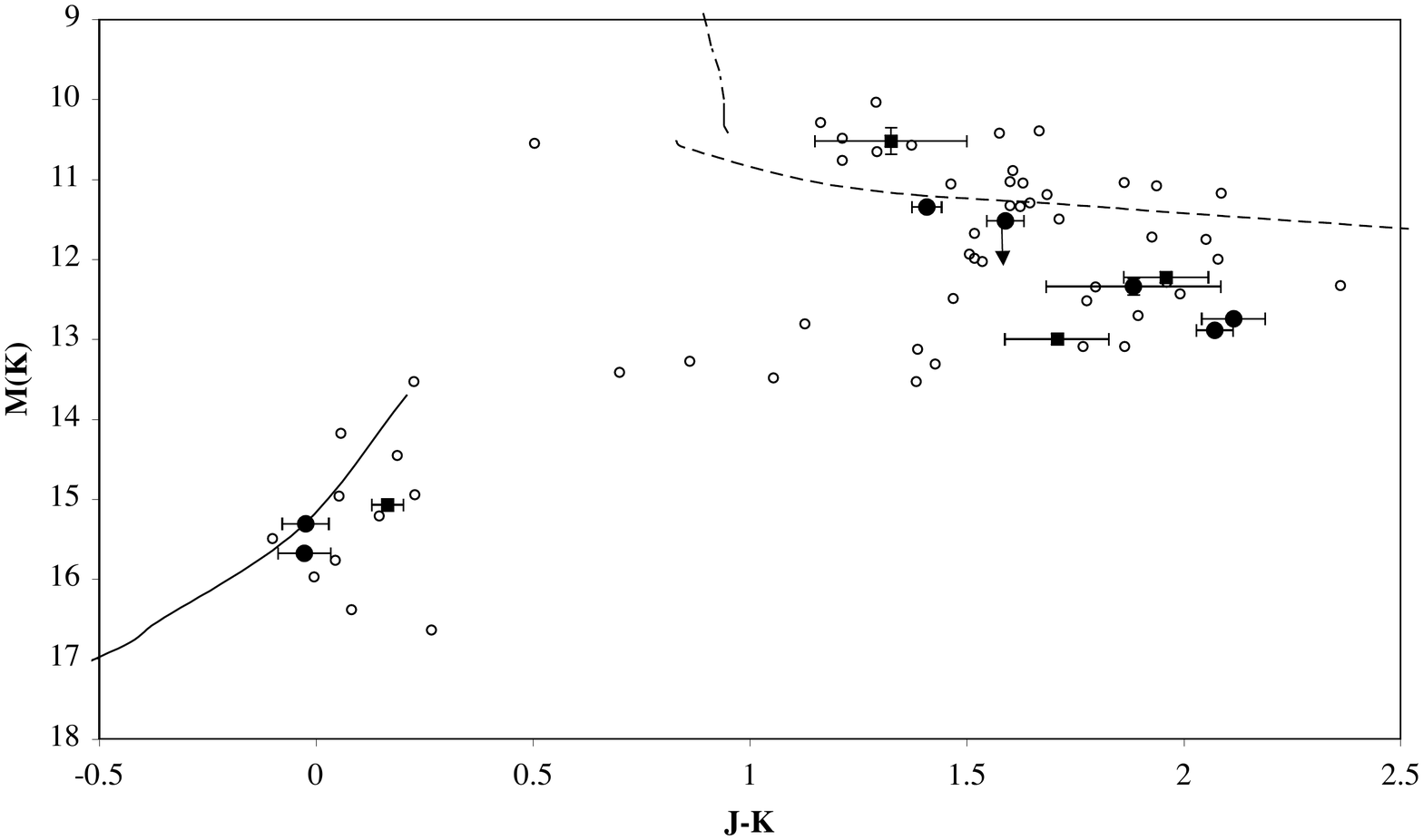}}
\vspace{-1.4cm}\caption{M$_{K}$ versus J-K diagram for stars in the L \& T dwarf list, with parallaxes. 0.5 Gyr isochrones are plotted (explicit 0.4 Gyr data was not available for condensed \& dusty models). Solid line: COND model. Dashed line: DUSTY model. Dot-dashed line: NEXTGEN model. Filled squares indicate the 4 UMSMG stars, and circles indicate the 7 HMG stars, identified as group members in this work. Open circles are field dwarfs not identified with a moving group.}
\label{fig:mk_jk}
\end{figure}

Fig.~\ref{fig:mk_jk} plots the $M_{K}$, $J-K$ colour magnitude diagram for both the UMSMG and the HMG members. Also shown are the 60 other L and T dwarfs from the archive with known parallaxes (2M1228-15 A and B are plotted). These show a rather scattered distribution which is to be expected since they presumably have a range of ages. In addition we have plotted the 500 Myr DUSTY model of \citet{Cha:2000} and the same age COND models (\citet{Bar:2003}). We draw the following conclusions:-

\begin{figure}
\vspace{0cm}
\hspace{-0.7cm}
\resizebox{!}{6.9cm}
{\includegraphics{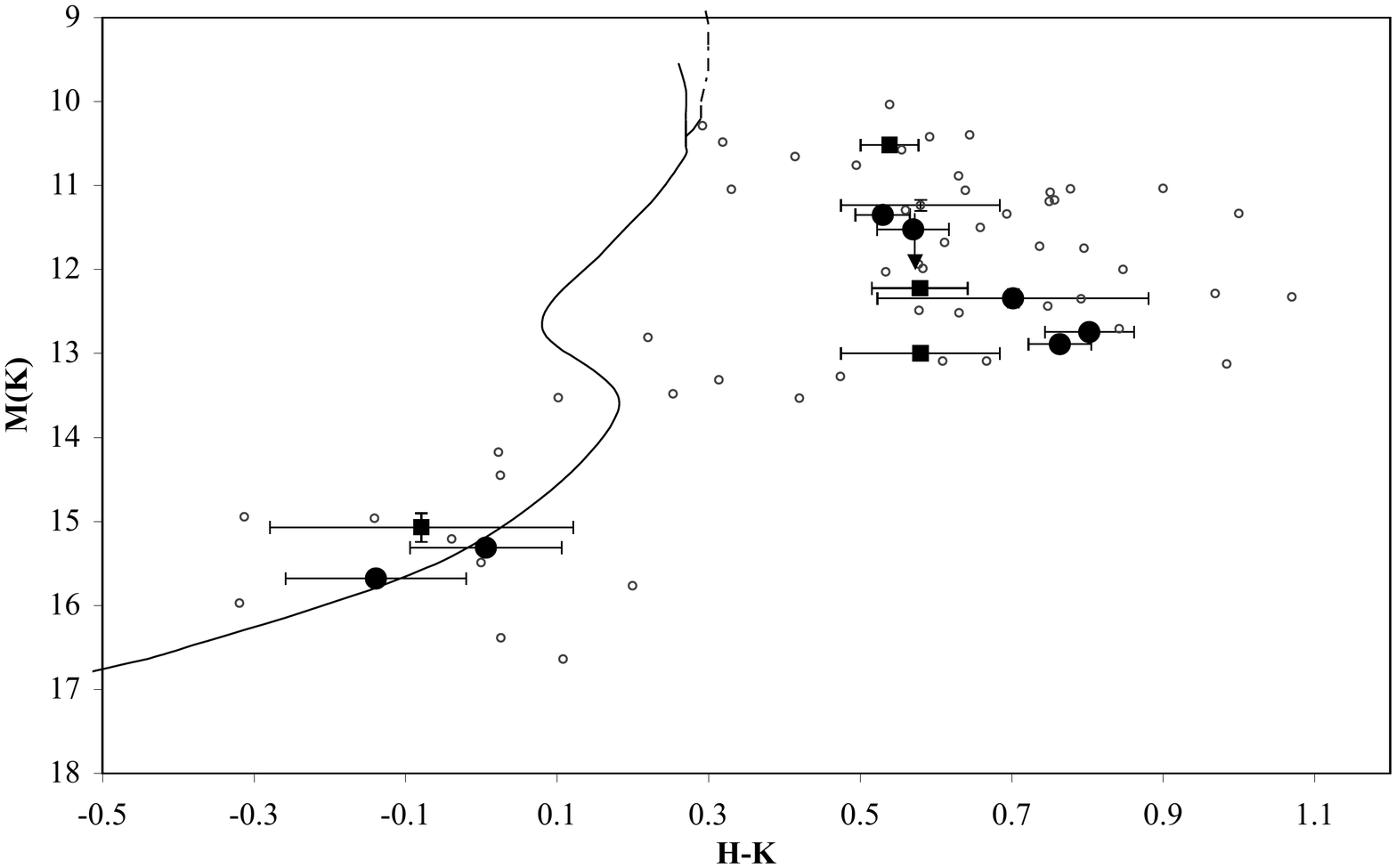}}
\vspace{-1.4cm}\caption{M$_{K}$ versus H-K diagram for stars in the L \& T dwarf list with parallaxes. 0.5 Gyr isochrones are plotted (explicit 0.4 Gyr data was not available for the condensed model). Solid line: COND model. Dot-dashed line: NEXTGEN model. The DUSTY model does not give H magnitudes and is therefore not shown. Filled squares indicate the 4 stars identified in this work as members of the Ursa Major-Sirius moving group, filled circles represent the 7 HMG members. Open circles are other field dwarfs.}
\label{fig:mk_hk}
\end{figure}

\begin{enumerate}
\item The 5 HMG L dwarfs sit on a very tight sequence. This suggests that coeval L dwarfs, unlike field L dwarfs, form a well defined sequence. We presume that the scattered nature of the field L dwarfs is therefore caused by their variety of ages or gravities, and possibly also metallicities.

\item Two of the UMSMG L dwarfs, if joined by a straight line, sit on a sequence that is nearly parallel and about 0.4 mag above the HMG sequence. This is as expected since the UMSMG is younger than the HMG. Any L dwarf sequence must have a turning point beyond which $J-K$ is decreasing towards the T dwarfs. The HMG sequence reaches $(J-K)_{2MASS} \sim 2.0$, close to the maximum $(J-K)_{MKO} \sim 2.0$~\citep{Leg:2000} and must therefore be close to or have reached this turning point. The UMSMG L8 dwarf 2M1523+3014 must be on the blueward arm of the L dwarf sequence. The HMG L6 dwarf 2M0825+2115 might also just be on the blueward arm of the sequence.

\item None of the L dwarfs except the uncertain 2M0205 and discarded 2M1228 appear to be low mass ratio binaries, otherwise they would sit high on the sequence. The only possible exception might be 2M1523+3014 which is alone on the blueward sequence; however, given its locus on the lower envelope of the field stars we think it is unlikely to be a binary. A possible reason for the lack of near-equal mass binaries is that these are of course more massive entities and therefore less likely to have been ejected from their parent clusters.

\item The 3 T dwarfs appear to form a short sequence that corresponds to an approximate 500 Myr isochrone. At its red end this lies below the COND model~\cite{Bar:2003}.

We have found 11 out of 70 field dwarfs to be members of the UMSMG and HMG. This is a high percentage, nearly 16\%. This of course implies that if we had proper motions and parallaxes for all 459 dwarfs in the dwarf archive, we would have found $459\times 11 / 70 = 72$, 26 UMSMG and 46 HMG members. For the Hyades this would certainly increase the Mass Function (see~\citet{Giz:1999}), but not ridiculously so given that the Hyades cluster is known to be deficient in low mass stars and brown dwarfs. 

That a significant percentage of field dwarfs belong to local moving groups may be due to very old (age $>1$ Gyr) dwarfs being very cool, faint and therefore difficult to detect and thus under-represented in the dwarf archive.

\item It can be seen that the DUSTY 500 myr isochrone is not a good fit to either UMSMG (age 400 myr) or the HMG (age 625 myr). This is perhaps not suprising given the theoretical difficulties of modelling dusty atmospheres.~\citet{Bur:2006} do not give theoretical isochrones. The COND model for T dwarfs fits the HMG at $J-K \simeq 0$ very well, but is not so good for the somewhat redder UMSMG T dwarf.

\item Figure~\ref{fig:mk_hk} shows the $M_{K}$, $H-K$ colour magnitude diagram. Again both moving groups have a well defined sequence in the L regime, but the UMSMG is almost vertical, whereas the HMG covers 0.3 in $H-K$ colour. Of course, 0.3 is not a great range in colour and it is already well known (e.g.~\citet{Leg:2000}) that the field L dwarfs have a small range in $H-K$ colour. Given that both moving groups have a similar age and therefore similar gravities, the most likely explanation for the difference is metallicity. The Hyades has Fe/H = +0.13 relative to the sun, while the UMSMG has -0.08~\citet{Kin:2005}.~\citet{Bur:2006} have theoretical models of L and T dwarfs and do indeed predict that L dwarfs will be redder in $J-K$ with increased metallicity. Unfortunately they do not consider the $H-K$ dependence on metallicity.

\end{enumerate}


\section{Coevality}
So far we have implied that the membership of a moving group guarantees objects have the same age. This is not necessarily true. Moving groups may arise from a dispersing cluster or from a star formation event in a particular region of a molecular cloud; in either case, the members of the group will be coeval. Alternatively a moving group may be the consequence of a dynamical process where for example the galactic bar drives some resonance to produce a group of stars with a common velocity. In this case the stars will not be coeval (see, for example,~\citet{Deh:1998}). The Pleiades moving group, sometimes called the local group, and other groups have stars of differing ages~\citep{Che:1998,Che:1999}. Also, the core stars of the UMSMG clearly fit a good sequence in the HR diagram and are coeval~\citep{Kin:2003}, but Sirius which, on dynamical grounds is a good member, probably does not have the correct age~\citep{Lie:2005}.

Resonance driven stars tend to favour particular $V$ (galactic azimuthal direction) velocities. These favoured velocities coincide with both the UMSMG and HMG $V$ velocities (see for example~\citet{Deh:1998} or~\citet{Sku:1999}). Thus there is a greater than random chance that stars will have proper motions pointing to the convergent point of these groups. This effect is rather difficult to quantify over the whole sky, so we have added an estimated factor of 4 in Section 2 for calculating the chance of a proper motion being directed towards the UMSMG or HMG's convergent point. However if an object has the UMSMG or HMG velocity by virtue of a dynamical resonance rather than from being a genuine member of the group, its moving group distance (see Section 2) will not be the same as its parallax distance and it will fail our second test. Perhaps fortuitously all our objects pass the second test.

Of course our third test, that the dwarfs fit a `sensible' sequence, should select coeval objects. However, we do not know exactly where this sequence is and the skeptic might argue that we have simply got the wrong sequence. Indeed with 2M1228-15 the first two tests produced an object that did not fit the UMSMG sequence (see above). On the positive side the 5 HMG L dwarfs do seem to form a very good sequence, and the 2 HMG T dwarfs have very similar absolute K magnitudes. The results for the UMSMG are not so compelling, but 2 L dwarfs lie on a line parallel and above the Hyades sequence as expected for a younger group. 

The results described in this paper should perhaps be treated with some caution. Nevertheless, the kinematic data exist and should be used. As more proper motion and parallax data become available these and other sequences may become better established.

Summarising, we can say that some moving group members are undoubtedly coeval, but membership of a moving group based on the criteria described in this paper does not guarantee coevality.

\section{Conclusion}
We find one T dwarf and 3 L dwarfs that belong to the UMSMG, whose age is 400 Myr. We find a further 2 T dwarfs and 5 L dwarfs, one rather dubious, members of the HMG. These stars provide preliminary empirical isochrones for these ages.

We plan to extend this technique to other moving groups. Only 70 of the 459 archived dwarfs have the proper motions and parallaxes needed to identify them with moving groups. We believe that many more L and T dwarfs could be identified with moving groups if more proper motions and parallaxes were available. This would allow them to be assigned an age, although it should be remembered that moving group ages can never be totally secure.



\section*{Acknowledgments}
This research has benefitted from the M, L, and T dwarf compendium housed at DwarfArchives.org and maintained by Chris Gelino, Davy Kirkpatrick, and Adam Burgasser. The research makes use of data products from the Two Micron All Sky Survey, which is a joint project of the University of Massachusetts and the Infrared Processing and Analysis Center/California Institute of Technology, funded by the National Aeronautics and Space Administration and the National Science Foundation. We are grateful to the referee, M. Bessel, for his helpful comments which improved this paper.

\label{lastpage}


\begin{thebibliography}{99}
\bibitem[\protect\citeauthoryear{Baraffe et al.}{1998}]{Bar:1998}Baraffe, I., Chabrier, G., Allard, F. \& Hauschildt, P.~H.: 1998, A\&A 337, 403
\bibitem[\protect\citeauthoryear{Baraffe et al.}{2002}]{Bar:2002}Baraffe, I., Chabrier, G., Allard, F. \& Hauschildt, P.~H.: 2002, A\&A 382, 563
\bibitem[\protect\citeauthoryear{Baraffe et al.}{2003}]{Bar:2003}Baraffe, I., Chabrier, G., Barman, T.~S., Allard, F. \& Hauschildt, P.~H.: 2003, A\&A 402, 701

\bibitem[\protect\citeauthoryear{Berger et al.}{2005}]{Ber:2005}{{Berger}, E., {Rutledge}, R.~E., {Reid}, I.~N., {Bildsten}, L., 
	{Gizis}, J.~E., {Liebert}, J., {Mart{\'{\i}}n}, E., 
	{Basri}, G., {Jayawardhana}, R., {Brandeker}, A., {Fleming}, T.~A., 
	{Johns-Krull}, C.~M., {Giampapa}, M.~S., {Hawley}, S.~L. \& 
	{Schmitt}, J.~H.~M.~M.}, 2005, ApJ, 627, 960

\bibitem[\protect\citeauthoryear{Bouy et al.}{2003}]{Bou:2003}{{Bouy}, H., {Brandner}, W., {Mart{\'{\i}}n}, E.~L., 
	{Delfosse}, X., {Allard}, F. \& {Basri}, G.}, 2003, AJ, 126, 1526

\bibitem[\protect\citeauthoryear{Bouy et al.}{2005}]{Bou:2005}{{Bouy}, H., {Mart{\'{\i}}n}, E.~L., {Brandner}, W. \& 
	{Bouvier}, J.}, 2005, AJ, 129, 511
	
\bibitem[\protect\citeauthoryear{Brandner et al.}{2004}]{Bra:2004}{{Brandner}, W., {Mart{\'{\i}}n}, E.~L., {Bouy}, H., 
	{K{\"o}hler}, R., {Delfosse}, X., {Basri}, G. \& {Andersen}, M.
	}, 1998, AA, 428, 205
	
\bibitem[\protect\citeauthoryear{Burgasser et al.}{2003}]{Burg:2003}{{Burgasser}, A.~J., {Kirkpatrick}, J.~D., {Reid}, I.~N., 
	{Brown}, M.~E., {Miskey}, C.~L. \& {Gizis}, J.~E.}, 2003, ApJ, 586, 512


\bibitem[\protect\citeauthoryear{Burgasser et al.}{2006}]{Burg:2006}{{Burgasser}, A.~J., {Burrows}, A. \& {Kirkpatrick}, J.~D.}, 2006, ApJ, 639, 1095

\bibitem[\protect\citeauthoryear{Burrows et al.}{1997}]{Bur:1997}{{Burrows}, A., {Marley}, M., {Hubbard}, W.~B., {Lunine}, J.~I., 
	{Guillot}, T., {Saumon}, D., {Freedman}, R., {Sudarsky}, D. \& 
	{Sharp}, C.}, 1997, ApJ, 491, 856

\bibitem[\protect\citeauthoryear{Burrows et al.}{2001}]{Bur:2001}{{Burrows}, A., {Hubbard}, W.~B., {Lunine}, J.~I. \& {Liebert}, J.}, 2001, {Reviews of Modern Physics}, 73, 719

\bibitem[\protect\citeauthoryear{Burrows et al.}{2006}]{Bur:2006}{Burrows, A., Sudarsky, D. \& Hubeny, I.}, 2006, ApJ, 640,1063

\bibitem[\protect\citeauthoryear{Carroll \& Ostlie}{1996}]{Car:1996}{{Carroll}, B.~W. and {Ostlie}, D.~A.}, 1996, ``{An introduction to modern astrophysics}'', {Reading, MA: Addison-Wesley}

\bibitem[\protect\citeauthoryear{Casewell, Jameson \& Dobbie}{2006}]{Cas:2006}{{Casewell}, S.~L. and {Jameson}, R.~F. and {Dobbie}, P.~D.},2006, AN, in press

\bibitem[\protect\citeauthoryear{Castellani et al.}{2002}]{Cas:2002}{{Castellani}, V. and {Degl'Innocenti}, S. and {Prada Moroni}, P.~G. and 
	{Tordiglione}, V.},2002, MNRAS, 334, 193
\bibitem[\protect\citeauthoryear{Chabrier et al.}{2000}]{Cha:2000} Chabrier, G., Baraffe, I., Allard, F. \& Hauschildt, P.~H., 2000, ApJ 542, 464
\bibitem[\protect\citeauthoryear{Chauvin et al.}{2004}]{Cha:2004}{Chauvin}, G., {Lagrange}, A.-M.,  {Dumas}, C., {Zuckerman}, B., {Mouillet}, D., {Song}, I., {Beuzit}, J.-L., {Lowrance}, P., 2004, AA, 425, L29

\bibitem[\protect\citeauthoryear{Chereul et al.}{1998}]{Che:1998}{{Chereul}, E., {Creze}, M. \& {Bienayme}, O.}, 1998, AA, 340, 384

\bibitem[\protect\citeauthoryear{Chereul et al.}{1999}]{Che:1999}{{Chereul}, E., {Creze}, M. \& {Bienayme}, O.}, 1999, AA Supp, 135, 5

\bibitem[\protect\citeauthoryear{Dehnen}{1998}]{Deh:1998}{{Dehnen}, W.}, 1998, AJ, 115, 2384

\bibitem[\protect\citeauthoryear{Epstein et al.}{2002}]{Eps:1997}{{Epstein, N. et al.}}, 1997, ESO Messenger, 87, 27

\bibitem[\protect\citeauthoryear{Gizis et al.}{1999}]{Giz:1999}{{Gizis}, J.~E., {Reid}, I.~N. \& {Monet}, D.~G.}, 1999, AJ, 118, 997


\bibitem[\protect\citeauthoryear{Jameson et al.}{2002}]{Jam:2002}{{Jameson}, R.~F., {Dobbie}, P.~D., {Hodgkin}, S.~T. and {Pinfield}, D.~J.},2002, MNRAS, 335, 853

\bibitem[\protect\citeauthoryear{King et al.}{2003}]{Kin:2003}{{King}, J.~R. and {Villarreal}, A.~R. and {Soderblom}, D.~R. and {Gulliver}, A.~F. and {Adelman}, S.~J.},2003, IAU Symposium 211,189

\bibitem[\protect\citeauthoryear{King and Schuler}{2005}]{Kin:2005}{{King}, J.~R. and {Schuler}, S.~C.},2005, PASP, 117, 911

\bibitem[\protect\citeauthoryear{Kirkpatrick et al.}{2001}]{Kir:2001}{{Kirkpatrick}, J.~D., {Dahn}, C.~C., {Monet}, D.~G., {Reid}, I.~N., {Gizis}, J.~E., {Liebert}, J. \& {Burgasser}, A.~J.}, 2001, AJ, 121, 3235


\bibitem[\protect\citeauthoryear{Kirkpatrick}{2003}]{Kir:2003} {Kirkpatrick, J.D.}, 2003, {IAU Symposium}, ed. {{Mart{\'{\i}}n}, E.}, 211, 189

\bibitem[\protect\citeauthoryear{Koerner et al.}{1999}]{Koe:1999}{{Koerner}, D.~W., {Kirkpatrick}, J.~D., {McElwain}, M.~W. \& 
	{Bonaventura}, N.~R.}, 1999, ApJ, 526, L25

\bibitem[\protect\citeauthoryear{Leggett et al.}{2000}]{Leg:2000}{{Leggett}, S.~K. et al.}, 2000, ApJ, 536, L35
	
\bibitem[\protect\citeauthoryear{Liebert et al.}{2005}]{Lie:2005} {{Liebert}, J. and {Young}, P.~A. and {Arnett}, D. and {Holberg}, J.~B. and 
	{Williams}, K.~A.},2005, ApJ, 630, L69

\bibitem[\protect\citeauthoryear{Madsen, Dravens and Lindegren}{2002}]{Mad:2002}{{Madsen}, S. and {Dravins}, D. and {Lindegren}, L.},2002, A\&A, 381, 446	
	
\bibitem[\protect\citeauthoryear{Mamajek}{2005}]{Mam:2005}{{Mamajek}, E.~E.},2005, ApJ, 634, 1385
\bibitem[\protect\citeauthoryear{McCook \& Sion}{1999}]{Mcc:1999} {{McCook}, G.~P. and {Sion}, E.~M.},1999, ApJ, 121, 1
\bibitem[\protect\citeauthoryear{Moraux et al}{2003}]{Mor:2003}{{Moraux}, E., {Bouvier}, J., {Stauffer}, J.~R. and {Cuillandre}, J.-C.}, 2003, AA, 400, 891

\bibitem[\protect\citeauthoryear{Odenkirchen, Soubiran \& Colin}{1998}]{Ode:1998}{{Odenkirchen}, M. and {Soubiran}, C. and {Colin}, J.},1998, New Astron., 3, 583
\bibitem[\protect\citeauthoryear{Perryman et al.}{1998}]{Per:1998} {{Perryman}, M.~A.~C. and {Brown}, A.~G.~A. and {Lebreton}, Y. and 
	{Gomez}, A. and {Turon}, C. and {de Strobel}, G.~C. and {Mermilliod}, J.~C. and 
	{Robichon}, N. and {Kovalevsky}, J. and {Crifo}, F.},1998, AA, 331, 81

\bibitem[\protect\citeauthoryear{Schweitzer et al.}{2001}]{Sch:2001}{{Schweitzer}, A. and {Gizis}, J.~E. and {Hauschildt}, P.~H. and 
	{Allard}, F. and {Reid}, I.~N.},2001, ApJ, 555, 368

\bibitem[\protect\citeauthoryear{Skrutskie et al.}{1997}]{Skr:1997} {Skrutskie, M.F. et al.},1997,``{The Impact of Large Scale Near-IR Sky Surveys}'', eds. F. Garzon et al. (Kluwer, Netherlands), 25

\bibitem[\protect\citeauthoryear{Skuljan et al.}{1999}]{Sku:1999}{{Skuljan}, J., {Hearnshaw}, J.~B. \& {Cottrell}, P.~L.}, 1999, MNRAS, 308, 731

\bibitem[\protect\citeauthoryear{Soderblom and Mayor}{1993}]{Sod:1993}{{Soderblom}, D.~R. and {Mayor}, M.},1993, ApJ, 105, 226


\bibitem[\protect\citeauthoryear{Vrba et al.}{2004}]{Vrb:2004}{{Vrba}, F.~J., {Henden}, A.~A., {Luginbuhl}, C.~B., 
	{Guetter}, H.~H., {Munn}, J.~A., {Canzian}, B., {Burgasser}, A.~J., 
	{Kirkpatrick}, J.~D., {Fan}, X., {Geballe}, T.~R., 
	{Golimowski}, D.~A., {Knapp}, G.~R., {Leggett}, S.~K., 
	{Schneider}, D.~P. \& {Brinkmann}, J.}, 2004, AJ, 127, 2948

\bibitem[\protect\citeauthoryear{York et al.}{2003}]{Yor:2000}{{York}, D.G. et al.}, 2000, AJ, 120, 1579

\bibitem[\protect\citeauthoryear{Zuckermann \& Song}{2004}]{Zuc:2004} {{Zuckerman}, B. and {Song}, I.},2004, ARAA, 42, 685	

\end{thebibliography}
\end{document}